\documentclass[preprint,12pt]{elsarticle}
\usepackage{booktabs,makecell, multirow, tabularx}
\usepackage{amssymb}
\usepackage{amsmath}
\newtheorem{definition}{Definition}
\newtheorem{corollary}{Corollary}
\newtheorem{theorem}{Theorem}
\newtheorem{lemma}[theorem]{Lemma}
\newdefinition{rmk}{Remark}

\journal{}

\begin{document}
\begin{sloppypar}
\begin{frontmatter}


\title{Sliding Block Martingale based Multi-hop Delay QoS Analysis}

\author[1]{Yuchao Dang}
\ead{dangyc23@mails.jlu.edu.cn}
\ead[url]{dangyc23@mails.jlu.edu.cn}

%\address[1]{, Street 129, 1043 NX Amsterdam, The Netherlands}
\affiliation[1]{organization={College of Communication Engineering, Jilin University},
                addressline={5372 South Lake Road}, 
                city={Changchun},
%               citysep={}, % Uncomment if no comma needed between city and postcode
                postcode={130012}, 
                state={Jilin},
                country={China}}

\author[1]{Xuefen Chi\corref{cor1}}
% \cormark[1]
\ead{chixf@jlu.edu.cn}
\ead[url]{chixf@jlu.edu.cn}

\cortext[cor1]{Corresponding author}

\nonumnote{This work was supported by the Jilin provincial Scientific and Technological Development Program under [grant numbers 20230101063JC].
  }

%\author[1]{Yuchao Dang}
%\ead{dangyc23@mails.jlu.edu.cn}
%\ead[url]{dangyc23@mails.jlu.edu.cn}
% \credit{Conceptualization of this study, Methodology, Software, Data curation, Writing - Original draft preparation}

%\affiliation[1]{organization={College of Communication Engineering, Jilin University},
%                addressline={5372 South Lake Road}, 
 %               city={Changchun},
%               citysep={}, % Uncomment if no comma needed between city and postcode
 %               postcode={130012}, 
 %               state={Jilin},
 %               country={China}}
%\author[1]{Xuefen Chi}[style=chinese]
%\cormark[1]
%\ead{chixf@jlu.edu.cn}
%\ead[url]{chixf@jlu.edu.cn}

%% Abstract
\begin{abstract}
With the growing density of wireless networks and demand for multi-hop transmissions, precise delay Quality of Service (QoS) analysis has become a critical challenge. This paper introduces a multi-hop delay QoS analysis framework based on the sliding block martingale, addressing the loose boundary issue of prior methods that rely on service process martingales and min-plus transformations. By constructing a sliding block martingale with a window, we capture both long-term trends and short-term fluctuations in the backlog, eliminating the reliance on the generalized incremental property. The framework redefines delay unreliability events using cascading attributes, deriving a more compact Delay Unreliability Probability Boundary (DUPB). To improve the efficiency of solving the key parameter $\theta$, we propose a Micrometric Intervals based Supermartingale Upcrossing Estimate Theorem, quantifying the upper bound of event occurrence frequency to constrain the solution space of $\theta$. Simulations based on the 3GPP UMa/UMi channel model validate the framework's effectiveness. Results show that in 2-7 hop scenarios, the maximum deviation between theoretical boundaries and Monte Carlo simulations is $4.116 \times 10^{-5}$, with a lower RMSE than existing methods. Iteration count and CPU time for solving $\theta$ are reduced by $59\%-72\%$ and $60.6\%-70.5\%$, respectively, improving analysis efficiency. Furthermore, the derived minimum service rate for multi-hop queues offers a valuable reference for resource allocation. The framework demonstrates high accuracy, scalability, and practicality in complex multi-hop networks.
\end{abstract}

\begin{graphicalabstract}
\end{graphicalabstract}

\begin{highlights}
\item Sliding Block Martingale for Fine Delay Analysis: A novel sliding block martingale framework is proposed, capturing both short-term fluctuations and long-term trends of backlog processes. This eliminates reliance on the generalized incremental property of service processes, enabling more precise statistical modeling of multi-hop delays.  
\item Tighter Delay Unreliability Probability Bound (DUPB): By redefining multi-hop delay events via concatenation properties and integrating sliding block martingales, a compact DUPB is derived. Theoretical bounds show a maximum deviation of \(4.116 \times 10^{-5}\) from Monte Carlo simulations in 2-7 hop scenarios, outperforming existing methods.  
\item Efficient Parameter Optimization: A Micrometric Intervals-based Supermartingale Upcrossing Theorem quantifies delay event occurrence rates, narrowing the solution space for the critical decay factor \(\theta\). This reduces iteration counts by $59\%-72\%$ and CPU time by $60.6\%-70.5\%$, significantly enhancing QoS analysis efficiency.  
\item Practical Resource Allocation Metrics: An explicit minimum service rate for multi-hop queues is derived, serving as a lower bound for per-hop transmission rates. Validated under varying traffic demands, this metric guides bandwidth and power adjustments while ensuring QoS guarantees.  
\item Scalability and Robustness: Extensive simulations under 3GPP UMa/UMi channel models confirm the framework’s scalability. The DUPB maintains compactness even in 7-hop scenarios, avoiding tail divergence observed in prior methods.  
\end{highlights}

%% Keywords

\end{frontmatter}

\section{Introduction}
% On the one hand, the lack of delay analysis in resource allocation may result in the failure of intelligent algorithms to adequately consider delay QoS guarantee and user experience during optimization. On the other hand, complex and unprocessed network state information may hinder the rapid convergence of optimal strategies. Therefore, conducting multi-hop delay QoS analysis is essential. By utilizing delay QoS analysis, network state information can be quantified into more explicit QoS guarantee metrics, aiding in faster and better convergence to the optimal strategy during iterative optimization.
% Martingale, an observational stochastic process based on a sub-$\sigma$ algebra (generated by the historical state of the target stochastic process) \cite{r1}. Subsequently, conditional expectation is employed to construct equations or inequalities between future and historical states, thereby enabling the prediction of the state of the stochastic process.

The global traffic is experiencing rapid growth, leading to an increasing demand for network capacity \cite{n2}. The densification of wireless networks is an inevitable trend. Traditional fiber-optic backhaul deployments account for more than $80\%$ of network construction costs. Therefore, reducing fiber dependence and optimizing backhaul costs have become crucial. Integrated Access and Backhaul (IAB) networks have emerged as a widely discussed solution due to their flexibility and low-cost advantages of wireless backhaul \cite{n1}. In recent years, IAB networks have achieved high-throughput communication by integrating millimeter-wave \cite{n3}, utilizing its abundant spectrum resources. However, unique challenges of millimeter waves, such as severe path loss, susceptibility to blockage, and interference caused by dense deployment, pose significant difficulties in the effective utilization of spectrum resources. In IAB networks, the delay analysis becomes particularly complex due to the multi-layer base station structure and the demand for multi-hop transmission. Cascaded queue delays in multi-hop links directly affect network throughput and user experience. To meet delay constraints, many traditional studies have employed excessive spectrum allocation to mitigate interference. However, this approach often results in wasted spectrum resources, potentially as high as $20\%-30\%$.

Currently, many resource allocation strategies leveraging intelligent technologies, such as Deep Reinforcement Learning (DRL), have been proposed to optimize resource distribution and ensure Quality of Service (QoS). For example, in \cite{n4}, the authors collect real-time network state information and use DRL to map this network state as input and resource strategies as output, aiming to improve resource allocation while guaranteeing QoS. Similarly, Cheng et al. utilize agent iterations to optimize resource strategy performance, but their approach directly using all the network information as states to iteratively refine the agent's policy \cite{n5}. In \cite{n6}, Wang et al. propose a multi-agent collaborative framework that combines the Multi-Head Self-Attention (MHSA) mechanism and Long Short-Term Memory (LSTM) network to capture system state changes, thus improving throughput and transmission delay. Jia et al. design a task offloading and resource allocation strategy based on distributed Deep-Q-Network, making offloading decisions based on locally observed state, and optimizing both computational and communication resources to maximize data processing efficiency \cite{n7}. Zhao et al. present an adaptive DRL-based joint resource allocation scheme, which dynamically adjusts resource strategies using both current states and historical data \cite{n8}.

While these studies address delay constraints, they often lack a rigorous QoS analysis. Researchers typically use network data as states for DRL algorithms and rely on the agent's iterative optimization of allocation strategies. However, this approach overlooks the critical impact of delay QoS on user experience. Without delay analysis, intelligent algorithms may fail to properly account for delay constraints, resulting in suboptimal resource allocation and degraded performance. Moreover, the complex and unprocessed nature of the network state information may hinder the rapid convergence of optimal strategies. Therefore, conducting multi-hop delay QoS analysis is crucial. By incorporating delay QoS analysis, network state information can be quantized into explicit QoS metrics, facilitating faster and more accurate convergence of the optimal resource allocation strategy during the iterative optimization process. This approach ensures that delay constraints are effectively considered, enhancing both system efficiency and user experience.

Accurate delay QoS analysis results are an excellent guide for resource allocation. Martingale is an observational stochastic process based on a sub-$\sigma$ algebra (generated by the historical state of the target stochastic process) \cite{r1}. Subsequently, conditional expectation is employed to construct equations or inequalities between future and historical states, thus facilitating the prediction of the stochastic process' state. In \cite{r2}, service (arrival) martingales are constructed by exploiting the generalized incremental property of the service (arrival) process and the supermartingale properties of the queue. As time progresses, the martingales gradually approaches the 'true' statistical features of process. This construction exploits the advantage of directly observe the stochastic process through the martingale. Furthermore, \cite{r2} demonstrates that QoS analysis based on service martingale theory provides more accurate results (i.e., Delay Unreliability Probability Bound (DUPB) ) compared to traditional methods. In \cite{rr24}, the authors utilized service martingale theory alongside stochastic network calculus to conduct a delay analysis, thereby improving the accuracy of QoS analysis. Reference \cite{rr7_1} presents a super-martingale-based mathematical model that links service processes with arrival flows to derive probability bounds for end-to-end delays. Reference \cite{rr7} examined the impact of multi-hop transmission on delay in Vehicular Ad Hoc Networks (VANETs), employing service martingale theory to address stochastic traffic and dynamic channels. In \cite{rr6}, the authors analyzed delay in multi-core edge computing systems by combining service martingale theory with Lyapunov optimization. In \cite{rr9}, an analytical framework for end-to-end latency in multihop systems was developed, and derived bandwidth requirements based on DUPB. Reference \cite{rr10} presented a predictive two-timescale resource allocation scheme for video-on-demand services, utilizing a service martingale theory based estimation method. 

The data backlogged and cannot leave the queue in time, which generates delay. Thus, short-term changes in backlog provide a more intuitive measure of delay than the service process. However, in existing methods \cite{r2,rr24,rr7_1,rr7,rr6,rr9,rr10}, the service martingale processes are constructed based on service, and the fact that the backlog process is non-generalized incremental leads to the inability of this martingale process to capture the short-term characteristics of the backlog. Many delay analyses studies \cite{rr24,rr7,rr6,rr9,rr10} use the min-plus to transform delay unreliability events into events described by arrivals and services. However, min-plus expands the event sample space, affecting DUPB's compactness. Moreover, the decay factor $\theta$ in DUPB is critical, but existing queue's steady-state formulations fail to constrain $\theta$ \cite{rr24,rr7_1,rr7,rr6,rr9}, resulting in a large solution space and thus reducing the efficiency of delay QoS analysis. To address these challenges above, this work introduces a Sliding Block Martingale multi-hop delay QoS analysis framework, which fundamentally enhances the precision and efficiency of multi-hop delay QoS analysis. The contributions of this paper as follow:

\begin{itemize}
\item 
The Sliding Block Martingale for fine statistical features: Backlog is inherently related to delay and is directly observable, however, its non-generalized incremental nature prevents us from using traditional construction methods to form a martingale for the observed backlog process. To fully exploit the direct observation advantage of backlog and martingale, sliding block martingale with finer statistical features be constructed. The sliding block martingale preserves both short-term fluctuations and long-term trends while eliminating the dependence of previous martingale constructions on the generalized incremental property of the observed process.

\item
Tighter delay unreliability probability bound and reference metric for resource allocation: Since existing DUPBs are loosened by min-plus event transformations, we redefine multi-hop delay and delay unreliability events using the concatenation property. Combined with the sliding block martingale, the Multi-hop Delay Unreliability Probability Bound theorem is proposed, which derived a more compact DUPB, and this DUPB also has good scalability for multi-hop scenarios. Based on the properties of multi-hop queues and the derived theorem, an explicit expression is derived for a lower bound on the transmission rate of each hop of a multi-hop queue, which provides a reference metric for resource allocation.

\item 
Efficient Solution of Key Parameters in QoS Analysis: Since the construction method of the Sliding Block Martingale has changed, the previous queue's steady-state condition is no longer applicable. Therefore, we propose the Steady-State Condition based Sliding Block Martingale Lemma, which provides a new steady-state condition for queues in the context of sliding block martingales. To quickly and accurately obtain the parameter $\theta$, we introduce the Micrometric Intervals based Supermartingale Upcrossing Estimate Theorem, which quantifies the maximum occurrences of the supermartingale process within a given interval. By combining this theorem with the necessary event of delay unreliability, we derive the Maximum Rate of Occurrence of the Delay Unreliability Necessary Event Corollary. This corollary provides the maximum frequency at which the delay unreliability event can occur, thus helping to narrow the solution space for the key parameter $\theta$ in the DUPB, ultimately improving the efficiency of QoS analysis.

\end{itemize}
The framework is validated through extensive simulations under 3GPP UMa/UMi channel models, demonstrating tight alignment between theoretical bounds and empirical results across 2–7 hop scenarios. Additionally, the derived minimum service rate for QoS guarantees adapts effectively to varying traffic demands, offering practical insights for resource allocation. The remainder of the paper is organized as follows: Section 2 details the Sliding Block Martingale framework and theorems' derivation; Section 3 presents numerical validation; Section 4 concludes the work.

\section{The Sliding Block Martingale based Multi-hop Delay QoS Analysis Framework}
\begin{figure}[t]
\centering
\includegraphics[height=0.7in]{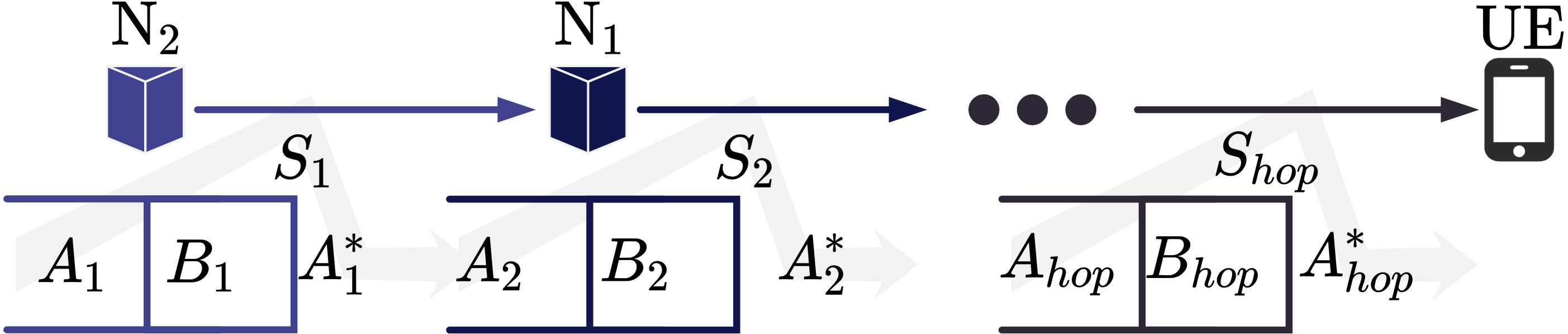}
\caption{Multi-hop Transmission}
\label{fig_0}
\end{figure}
In a single hop queue, $A\left(t\right):=\sum\limits_{i=1}^t{a\left(t\right)}$ is the arrival process, $S\left(t\right):=\sum\limits_{i=1}^t{s\left(t\right)}$ is the service process, and backlog process is expressed as $Q\left(t\right)=\sum\limits_{i=1}^t{q\left(t\right)}$. All data received by the UE comes from the CN, the downlink transmission from the CN to the UE, there are multiple hops, here we abstract each downlink transmitter into a node, using the symbols $N_i, i=1,2,\ldots$ to denote. As shown in Fig.~\ref{fig_0}, there are ${hop}$ hops between ${\rm N_1}$ and UE, the arrival and departure of the multi-hop tandem queue of UE are $A_1\left(t\right)$ and $A^*_{{hop}}\left(t\right)$, respectively. And the delay of UE is:
\begin{equation}
\label{eq_dbmqos_11_b1}
\begin{array}{llll}
W\left(t\right):=\inf\limits
\left\{\tau\ge 0:
\begin{matrix}
A_1\left(t\right) \le A^*_{{hop}}\left(t+\tau\right)
\end{matrix}
\right\}.\\
\end{array}
\end{equation}
Combined with $Q\left(t\right)=A\left(t\right)-A^*\left(t\right)$, we have
\begin{equation}
\label{eq_dbmqos_11_b2}
\begin{array}{cccc}
&\left\{A_1\left(t\right)\le A^*_{{hop}}\left(t+\tau\right)\right\}\\
\Leftrightarrow & \left\{Q_{{hop}}\left(t+\tau\right)\le {A_{{{hop}}-1}^*}\left(t+\tau\right)-A_1\left(t\right)\right\}\\
&\ldots \\
\Leftrightarrow & \left\{\sum_{i=1}^{{hop}}{Q_{i}\left(t+\tau \right)}\le A_1\left(t+\tau\right)-A_1\left(t\right) \right\}.\\
\end{array}
\end{equation}
Therefore, as depicted in (\ref{eq_dbmqos_11_b1}) and (\ref{eq_dbmqos_11_b2}), in $\left[t,t+W\left(t\right)\right]$, the data backlogged across all nodes equals the cumulative arrivals in ${\rm N_1}$.
Hence, $W\left(t\right)$ is redefined as:
\begin{equation}
\label{eq_dbmqos_11_b3}
\begin{array}{llll}
W\left(t\right):=\inf\limits
\left\{\tau\ge 0:
\begin{matrix}
\sum_{i=1}^{{hop}}{Q_{i}\left(t+\tau \right)}\le A_1\left(t+\tau\right)-A_1\left(t\right)
\end{matrix}
\right\},\\
\end{array}
\end{equation}

The instantaneous backlog rate is
\begin{equation}
\label{eq_dbmqos_1}
\begin{array}{lll}
q\left(t\right)=\left(a\left(t\right)-s\left(t\right)\right)\mathbf{1}_{\left\{Q\left(t\right)> 0\right\}}-Q\left(t-1\right)\mathbf{1}_{\left\{Q\left(t\right)= 0\right\}},
\end{array}
\end{equation}
when the queue is non-empty, the random process $\left\{q\left(t\right)\mid t\in\mathbb{N}^{+}\right\}$ is independent and stationary, i.e., it is feasible to construct a martingale based on $Q\left(t\right)$ in this state. 
The derivation of (\ref{eq_dbmqos_1}) and the proof of these properties are provided in \ref{apped:pfcmb}.
Then, inspired by \cite{r2}, the sub-$\sigma$ algebra generated by the historical information of $Q\left(t\right)$ enables the martingale to configure the statistical properties of $Q\left(t\right)$. Then, the backlog-martingale is defined as
\begin{definition}[Backlog-Martingale]
There exists a deterministic function $D_Q\left(\theta\right)=\frac{1}{\theta}\ln\mathbb{E}\begin{bmatrix}e^{\theta q\left(t\right)}\end{bmatrix}$
such that 
\begin{equation}
\label{eq_dbmqos_2}
\begin{array}{lll}
\left\{{rv}_{q(t)}:=e^{\theta\left(q\left(t\right)-D_Q\left(\theta\right)\right)}\mid t\in\mathbb{N}^{+}\right\}
\end{array}
\end{equation}
is a sequence of independent non-negative r.v. with $\mathbb{E}\left[{rv}_{q(t)}\right]=1$. Then, 
\begin{equation}
\label{eq_dbmqos_3}
\begin{array}{cll}
M_Q&:=\left\{M_Q\left(t,\theta\right)\mid t=1,2,\ldots\right\}\\
M_Q\left(t,\theta\right)&:=\prod\limits_{i=1}^t{rv}_{q(i)}=e^{\theta\left(Q\left(t\right)-tD_Q\left(\theta\right)\right)}
\end{array}
\end{equation}
is a martingale. The proof of correctness of constructing $M_Q$ is provided in the \ref{apped:pfcmb}.
\end{definition}
However, $Q\left(t\right)$ is non-generalized incremental and returns to zero after a certain time. $D_Q\left(\theta\right)$ records the average backlog rate, indicating insufficient information for computing DUPB.

To keep both long-term and short-term statistical information. Thus, the Sliding Block Martingale is defined as follows.

\begin{definition}[Sliding Block Martingale]
For the non-negative random process $\left\{x\left(t\right)\mid t\in\mathbb{N}^{+}\right\}$ is independent and stationary. Then, for $X\left(t\right):=\sum_{i=1}^{t}{x\left(t\right)}$, there exists a deterministic function 
\begin{equation}
\label{eq_dbmqos_4}
\begin{array}{cll}
D_{X}\left(\theta\right)=\frac{1}{\theta
 }\ln\mathbb{E}\begin{bmatrix}e^{\theta\left(\frac{X\left(t+{W}^b\right)-X\left(t\right)}{{W}^b}\right)}\end{bmatrix}
\end{array}
\end{equation}
such that $\left\{{rv}_{x(t)}:=e^{\theta\left(x\left(t\right)-D_{X}\left(\theta\right)\right)}\mid t\in\mathbb{N}^{+}\right\}$ is a sequence of independent non-negative r.v. with $\mathbb{E}\left[{rv}_{x(t)}\right]=1$. Then, 
\begin{equation}
\label{eq_dbmqos_5}
\begin{array}{cll}
{M_{SB}}\left(X,\theta,{W}^b\right)&:=\left\{{M_{SB}}\left(X\left(t\right),t,\theta,{W}^b\right)\mid t=1,2,\ldots\right\}\\
{M_{SB}}\left(X\left(t\right),t,\theta,{W}^b\right)&:=e^{\theta\left(X\left(t+{W}^b\right)-X\left(t\right)-{W}^bD_{X}\left(\theta\right)\right)}\\
\end{array}
\end{equation}
is martingale.
\end{definition}
The proof of correctness of constructing ${M_{SB}}$ is provided in the \ref{apped:pfcdm}.
The intuition of constructed Sliding Block Martingale: 
The constructed Sliding Block Martingale employs a fixed-size sliding window that incrementally advances over time to simultaneously preserve long-term and short-term statistical features. By aggregating the mean rate of the observed process over intervals $\left[i, i+W^b\right]$, the deterministic function $D_X(\theta)$ captures long-term statistical characteristics. Meanwhile, the condition $\mathbb{E}\left[e^{\theta\left(x(t)-D_X(\theta)\right)}\right]=1$ ensures the preservation of short-term dynamics through adaptive updates of $\theta$ based on $X(t)$, thereby eliminating reliance on the generalized incremental property of the stochastic process.

Based on the above, when the queue is non-empty, $\left\{q\left(t\right)\mid t\in\mathbb{N}^{+}\right\}$ and $\left\{a\left(t\right)\mid t\in\mathbb{N}^{+}\right\}$ both are non-negative independent and stationary random process, such that these random processes can be observed by Sliding Block Martingale. Then we have ${M_{SB}}\left(Q,\theta,{W}^b\right):=\left\{{M_{SB}}\left(Q_i\left(t\right),t,\theta,{W}^b\right)\mid t=1,2,\ldots\right\}$ and ${M_{SB}}\left(A,\theta,{W}^b\right):=\left\{{M_{SB}}\left(A\left(t\right),t,\theta,{W}^b\right)\mid t=1,2,\ldots\right\}$, both have the properties of Sliding Block Martingale. In the multi-hop transmission scenario shown in Fig.~\ref{fig_0}, with FIFO discipline, we present the Multi-hop Delay Unreliability Probability Bound theorem.
\begin{theorem}% []
\label{theorem1}
The ${hop}$-hop network with the backlogs $\left \{ Q_i\left(t\right)\mid i\in \left[1,{hop}\right] \right \} $ and the arrival $A_1\left(t\right)$ admit sliding block martingales ${M_{SB}}\left(Q_i,\theta,{W}^b\right)$ and ${M_{SB}}\left(A_1,\theta,{W}^b\right)$, respectively. Then, the multi-hop Delay Unreliability Probability Bound holds for any delay bound ${W}^b\ge 0$
\begin{equation}
\label{eq_dbmqos_14_b1}
\begin{array}{llll}
\mathbb{P}\left\{
W\left(t\right)\ge {W}^b
\right\}\le
\mathbb{E}\begin{bmatrix}{e^{\theta X_{msb}}}\end{bmatrix}e^{ \theta \sum\limits_{i=1}^{{hop}}{Q_{i}\left(t\right)}}.
\end{array}
\end{equation}
\end{theorem}
The proof is given in \ref{apped:pt2}.

% From Theorem \ref{theorem1}, it can be concluded that $\theta$ is a critical variable affecting the compactness of DUPB. Most studies solve $\theta$ based on the queue's steady-state condition \cite{rr24,rr7_1,rr7,rr6, rr8,rr9} (i.e., the mean arrival rate is less than or equal to the mean service rate), relying solely on an inequality, without numerical constraints on $\theta$, this introduces a challenge in solving $\theta$. The rate of an event's occurrence is greater than its probability. If the rate of occurrence of delay unreliability can be determined, an upper bound for $\theta$ can be derived from DUPB presented in Theorem \ref{theorem1}, providing a constraint interval for $\theta$. Then, we propose the Theorem \ref{theorem2} (Micrometric Intervals based Supermartingale Upcrossing Estimate Theorem).
From Theorem \ref{theorem1}, it can be concluded that $\theta$ is a critical variable affecting the compactness of DUPB. Most studies solve $\theta$ based on the queue's steady-state condition \cite{rr24,rr7_1,rr7,rr6,rr9} (i.e., the mean arrival rate is less than or equal to the mean service rate). 
Since the construction method of Sliding Block Martingale and its observation objects have been changed, we need to rederive the queueing steady state condition of Sliding Block Martingale based on arrival and backlog, so we propose the Steady-State Condition based on Sliding Block Martingale Lemma. 
\begin{lemma}
\label{lemma1}
The ${hop}$-hop network with the arrival $A_1\left(t\right)$ and backlogs $\left \{ Q_i\left(t\right)\mid i\in \left[1,{hop}\right] \right \} $ admit sliding block martingales ${M_{SB}}\left(Q,\theta,{W}^b\right)$ and ${M_{SB}}\left(A,\theta,{W}^b\right)$, respectively. Then, the multi-hop's queue steady-state condition can be expressed as: 
\begin{equation}
\small
\label{eq_dbmqos_lemma_1}
\begin{array}{lll}
\sum_{i=1}^{{hop}}{{W}^b{D_{ B_i}\left(\hat{\theta}\right)}}+\sum_{i=1}^{{hop}}Q_{i}\left(t \right)\le {{W}^b}D_{A_1}\left(\hat{\theta}\right). 
\end{array}
\end{equation}
% where $\sum_{i=1}^{{hop}}{{W}^b{D_{ B_i}\left(\hat{\theta}\right)}}+\sum_{i=1}^{{hop}}Q_{i}\left(t \right)\le {{W}^b}D_{A_1}\left(\hat{\theta}\right)$ is the queue's steady-state condition, with the proof provided in the \ref{apped:ssc}.
\end{lemma}
The proof of Lemma \ref{lemma1} is provided in the \ref{apped:ssc}.

% 无论是重新推导的稳态条件(\ref{eq_dbmqos_lemma_1})还是\cite{}使用的稳态条件, 它们都仍然存在一个挑战, 那就是如果只relies solely on inequality without numerical constraints on $\theta$, 那么$\theta$的求解效率是低的.
Whether it is the rederived steady-state condition (\ref{eq_dbmqos_lemma_1}) or the steady-state condition used in \cite{rr24,rr7_1,rr7,rr6,rr9}, both still face a challenge: if it solely relies on inequality without numerical constraints on $\theta$, the solving efficiency of $\theta$ is low. The rate of an event's occurrence is greater than its probability. If the rate of occurrence of delay unreliability can be determined, an upper bound for $\theta$ can be derived from DUPB presented in Theorem \ref{theorem1}, providing a constraint interval for $\theta$. Then, we propose the Theorem \ref{theorem2} (Micrometric Intervals based Supermartingale Upcrossing Estimate Theorem).
\begin{theorem}
\label{theorem2}
For the given random process $Y_T:=\left\{Y_t\mid t\in\left[1,T\right]\right\}$ stopped at $T$, which is a supermartingale and bounded in $\mathcal{L}^1$. Let ${O_T}\left[a\right]$ be the times of $Y_T$ greater than $a$
\begin{equation}
\label{eq_dbmqos_13_b1}
\begin{array}{lll}
{O_T}\left[a\right]:=\max\left\{
k\mid
\begin{matrix}\exists 0\le s_1<\ldots<s_k\le N\\
Y_{s_i} > a,\forall i\in\left[1,k\right]
\end{matrix}
\right\}
\end{array}
\end{equation}
% Divide $\left[a,\max\limits_{n\in\left[1,N\right]}\left\{X_n\right\}\right]$ into 
divide a series of sub-intervals $\left[{sb}_{l}^{i},{sb}_{h}^{i}\right]$, where 
\begin{equation}
\label{eq_dbmqos_13_b2}
\begin{array}{lll}
\left[{sb}_{l}^{i},{sb}_{h}^{i}\right],i\in\left[1,{sb}_{seg}\right]
,{sb}_{seg}=\left \lceil \frac{\max\left\{Y_T\right\}-a}{\delta } \right \rceil, \\
\begin{array}{lll}
{sb}_{l}^{i}&=\left\{
\begin{array}{lll}
a,&i=1\\
{sb}_{l}^{i-1}+\delta,&i\in\left[2,{sb}_{seg}\right],
\end{array}
\right .\\
{sb}_{h}^i&={sb}_{l}^{i}+\delta ,i\in\left[1,{sb}_{seg}\right],\delta =\mathbb{E}\left[Y_t-Y_{t-1}\right],\\
\end{array}
\end{array}
\end{equation}
and then ${O_T}\left[a\right]\le\sum_{i=1}^{{sb}_{seg}}{\frac{2\mathbb{E}\left[\left(Y_t-{sb}_{l}^{i}\right)^{-}\right]}{\delta }}$. The proof is detailed in \ref{apped:pt3}.
\end{theorem}

The core idea of Theorem \ref{theorem2}: Doob Upcrossing Lemma \cite{r1} provides the number of upcrossings for a supermartingale within a given real interval. By partitioning this interval into sufficiently small subintervals, the sum of upcrossing counts for each subinterval, multiplied by $2$, must greater than the rate of occurrence of the event where '$Y_T$' exceeds the lower bound of the real interval'. The factor of $2$ arises because, for a supermartingale, the conditional expectation decreases gradually, thus, an upcrossing within a subinterval implies a downcrossing within that same subinterval.

According to (\ref{eq_dbmqos_11_b3}), we have
\begin{equation}
\label{eq_dbmqos_13_b3}
\begin{array}{lll}
\left\{W\left(t\right)\ge {W}^b\right\}\subseteq 
\left\{\sum\limits_{i=1}^{{hop}}{Q_{i}\left(t+{W}^b \right)}\ge A_1\left(t+W^b\right)-A_1\left(t\right)\right\}.
\end{array}
\end{equation}

As can be seen, when the event on the left of (\ref{eq_dbmqos_13_b3}) occurs, the event on the right must necessarily occur. Therefore, we define $Y_t^{delay}:= \sum_{i=1}^{{hop}}Q_{i}\left(t+{W}^b \right) -A_1\left(t+{W}^b\right)+A_1\left(t\right)$ as the Delay Unreliability Necessary Event. Then, Corollary \ref{corollary4} (Maximum Rate of Occurrence of the Delay Unreliability Necessary Event) is proposed.
\begin{corollary}% []
\label{corollary4}
Delay Unreliability Necessary random process $Y_T^{delay}:=\left\{Y_t^{delay}\mid t\in\left[1,T\right]\right\}$ is defined as 
\begin{equation}
\label{eq_dbmqos_13_a1_1}
\begin{array}{lll}
Y_t^{delay}:= \sum_{i=1}^{{hop}}Q_{i}\left(t+{W}^b \right) -A_1\left(t+{W}^b\right)+A_1\left(t\right),\\
\end{array}
\end{equation}
and, the interval $\left[0,\underset{t\in\left[1,T\right]}{\arg\max}\left\{Y_t^{delay}\right\}\right]$ is partitioned as described in Theorem \ref{theorem2}.
Then, the Maximum Rate of Occurrence of the $\left\{W\left(t\right)\ge {W}^b\right\}$ is
\begin{equation}
\label{eq_dbmqos_13_a1_3}
\begin{array}{lll}
{mr}
&=\frac
{\sum_{i=1}^{{sb}_{seg}}{2\mathbb{E}\left[\left(Y_t^{delay}-{sb}_{l}^{i}\right)^{-}\right]}}{\delta T}.
\end{array}
\end{equation}
\end{corollary}
The proof is detailed in \ref{apped:c1}.

It is evident that the occurrence of event cannot be less than its probability, i.e., $\mathbb{P}\left\{W\left(t\right)\ge {W}^b\right\}\le{mr}$. 
Thus, based on Corollary \ref{corollary4} and Theorem \ref{theorem2}, we have, 
\begin{equation}
\label{eq_ptheta_1}
\begin{array}{lll}
{mr}
\ge \mathbb{E}\begin{bmatrix}{e^{\theta X_{msb}}}\end{bmatrix}e^{ \theta \sum_{i=1}^{{hop}}{Q_{i}\left(t\right)}}
\overset{a}{\ge} 
e^{\theta \mathbb{E}\begin{bmatrix}{{ X_{msb}}}\end{bmatrix}+\sum_{i=1}^{{hop}}{Q_{i}\left(t\right)}},
\end{array}
\end{equation}
where the step $a$ is based on the Jensen's inequality.
Then, $\theta \leq \frac{\ln(mr)}{\mathbb{E}[X_{msb}] + \sum_{i=1}^{hop}{Q_{i}(t)}}$. 

% $\theta$的区间约束给定之后, 另一个关键就是用构造的观测鞅过程来描述队列稳态条件 (均值服务不低于均值到达). 因此, 我们提出了如下Lemma.
% After the interval constraint of $\theta$ is specified, 

Based on the Lemma \ref{lemma1} and Corollary \ref{corollary4}, $\theta$ is solved by
\begin{equation}
\small
\label{eq_dbmqos_13_1}
\begin{array}{lll}
\theta=\sup\left\{\hat{\theta}\mid \begin{array}{ccc}
\hat{\theta} \leq \frac{\ln(mr)}{\mathbb{E}[X_{msb}] + \sum_{i=1}^{hop}{B_{i}(t)}},\\
\sum_{i=1}^{{hop}}{{W}^b{D_{ B_i}\left(\hat{\theta}\right)}}+\sum_{i=1}^{{hop}}Q_{i}\left(t \right)\le {{W}^b}D_{A_1}\left(\hat{\theta}\right)
\end{array}\right\},
\end{array}
\end{equation}

% 时延QoS分析的意义在于, 将采集的队列信息(到达, 服务和积压)和用户的业务QoS保障需求转化为可以为资源分配提供参考的信息. 基于定理2, 并且配置了QoS保障需求($W^b$ and $\varepsilon$), 我们得到了等式
The significance of delay QoS analysis lies in transforming the collected queue information (arrival, service, and backlog) and the user's service QoS requirements into reference information for resource allocation. Based on Theorem \ref{theorem2}, with $W^b$ and $\varepsilon$ configured and ${hop}$ known, then
\begin{equation}
\label{eq_dbmqos_16}
\begin{array}{lll}
\varepsilon= \mathbb{E}\begin{bmatrix}{e^{\theta X_{msb}}}\end{bmatrix}e^{ \theta \sum\limits_{i=1}^{{hop}}{Q_{i}\left(t\right)}}
\Rightarrow\frac{1}{\theta}\ln\left(\frac{\varepsilon_u}{\mathbb{E}\begin{bmatrix}{e^{\theta X_{msb}}}\end{bmatrix}}\right)={  \sum\limits_{i=1}^{{hop}}{Q_{i}\left(t\right)}}.\\
\end{array}
\end{equation}
Firstly, based on the Concatenation Property, the multi-hop tandem queue of the UE can be concatenated as a 1-hop queue \cite{r39}. To guarantee QoS for this 1-hop queue, the backlog $\sum_{i=1}^{{hop}}{Q_i\left(t\right)}$ must be cleared within ${W}^b$. If each subgroup within the multi-hop link group of the the UE is configured with the same service rate \( \sum_{i=1}^{hop} \frac{Q_i(t)}{W^b} \), then theoretically, this QoS of the the UE can be guaranteed.
Consequently, with the given QoS guarantee configuration, the minimum service rate $C$ the UE's multi-hop link is given by:
\begin{equation}
\label{eq_4_18a}
\begin{array}{l}
C=\frac{1}{\theta {W}^b}\ln\begin{pmatrix}\frac{\varepsilon}{\mathbb{E}\begin{bmatrix}{e^{\theta X_{msb}}}\end{bmatrix}}\end{pmatrix}.
\end{array}
\end{equation}
% 最小服务速率提供了一个具体的数值, 这个数值必然大于多跳队列中第一跳的到达过程的平均速率, 这也就为每一跳的传输速率提供了一个下限, 然后, 每一跳的信道最大可达速率应当高于这一下限. 又因为信道最大可达速率由信道传输SINR和带宽共同决定, 因此, 最小服务速率为功率调节和带宽调节提供了一个共同参考指标
The minimum service rate provides a specific value, which is guaranteed to be greater than the average arrival rate of the first hop in a multi-hop queue, thereby setting a lower bound for the transmission rate at each hop. The maximum achievable rate of the channel at each hop should exceed this lower bound. Furthermore, the maximum achievable rate of the channel is determined by both the channel's SINR and bandwidth. Therefore, the minimum service rate serves as a common reference metric for power and bandwidth adjustments.

\section{Simulation and Numerical Analysis}
\begin{figure}[ht]
\centering
\includegraphics[height=1.6in]{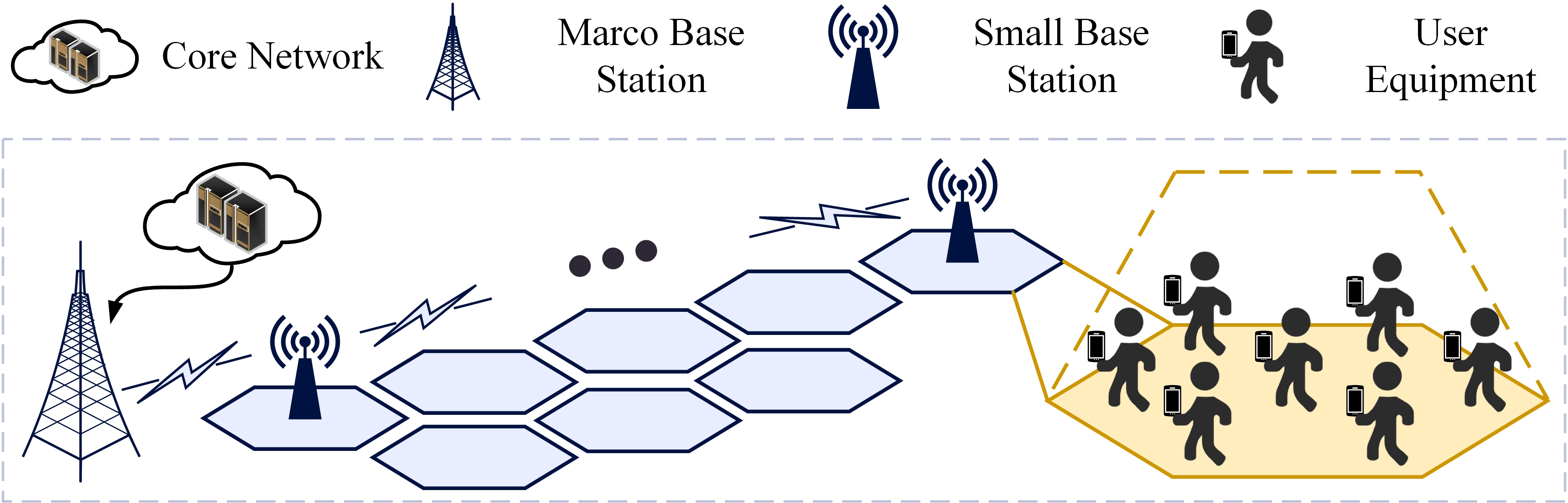}
\caption{Multi-hop Transmission, Uma and Umi}
\label{fig_1}
\end{figure}

To validate the theoretical framework, we conducted comprehensive simulations under multi-hop transmission scenarios in Fig.~\ref{fig_1}. The experimental setup, methodology, and comparative analysis are detailed as follows.

The wireless channel is modeled using the 3GPP TR 38.901 UMa (Urban Macrocell) and UMi (Urban Microcell) propagation models \cite{r5}. Path loss is calculated via the line-of-sight (LOS) and non-line-of-sight (NLOS) equations in Section 7.4.1 of the standard. Specifically, the UMa model adopts a carrier frequency of $28GHz$, a base station (BS) height of $25m$, and an inter-site distance of $500m$, while the UMi model assumes a BS height of 10 m, a street width of $20m$, and a building height of $20m$. Shadowing follows a log-normal distribution with standard deviations of $8.2dB$ (UMa) and $7.8dB$ (UMi). For the traffic model, the strength of the arrival process is determined by the generalized stochastically bounded bursty traffic \cite{r14}, and the distribution of user requests admits Poisson distribution.The service process was governed by channel capacity derived from the instantaneous Signal-to-Interference-plus-Noise Ratio (SINR), the formulation of SINR is given by \cite{r4}. Tandem queues with $2$ to $7$ hops were analyzed. Each node employed FIFO scheduling. 

Monte Carlo simulations were performed with $10^{7}$ independent realizations of the arrival process. For each realization, the end-to-end delay $W\left(t\right)$ was computed via (\ref{eq_dbmqos_11_b3}), and the empirical delay unreliability probability $\mathbb{P}\left\{W\left(t\right)\ge {W}^b\right\}$ was obtained by counting the ratio of events exceeding $W^b$.

\begin{figure}[h]
\centering
\includegraphics[height=2in]{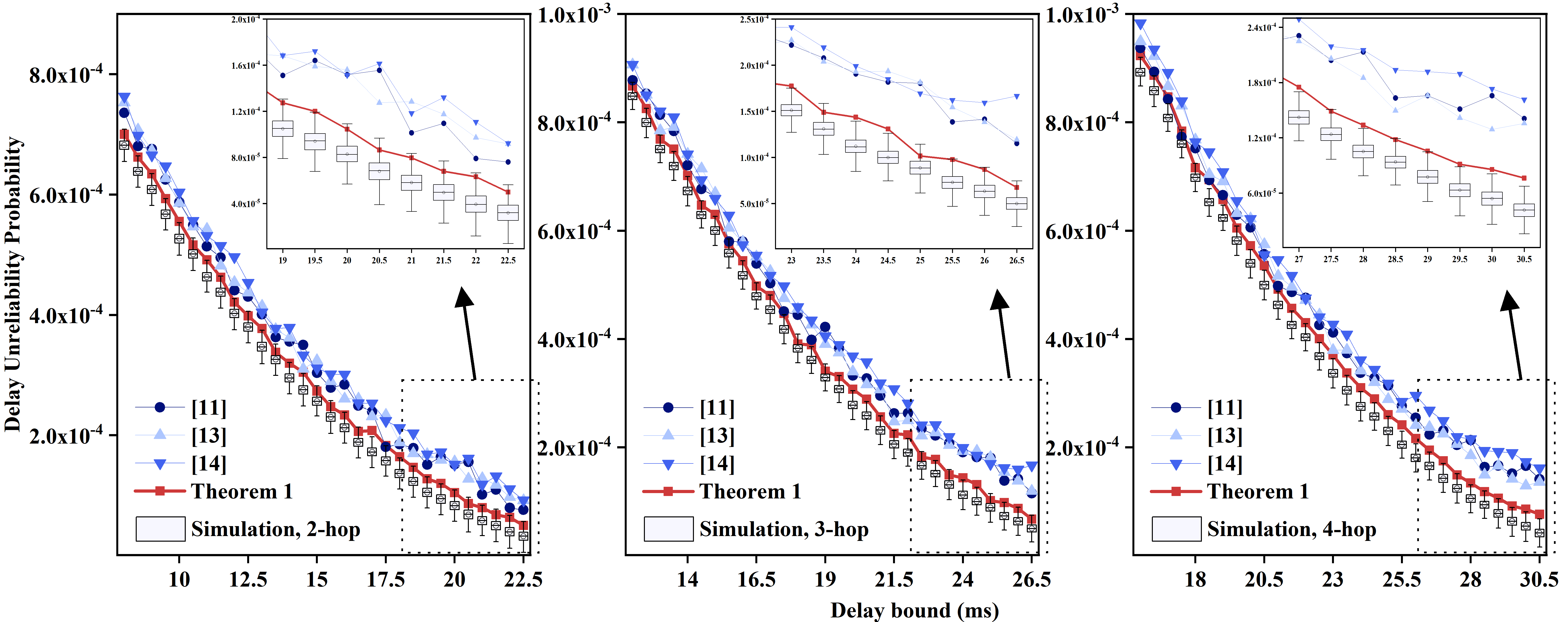}
\caption{Comparison of Delay unreliability probability bound compactness for Theorem \ref{theorem1}, \cite{rr24}, \cite{rr7}, \cite{rr6}. In multi-hop scenarios of 2 to 4 hops.}
\label{fig_2}
\end{figure}

\begin{figure}[h]
\centering
\includegraphics[height=2in]{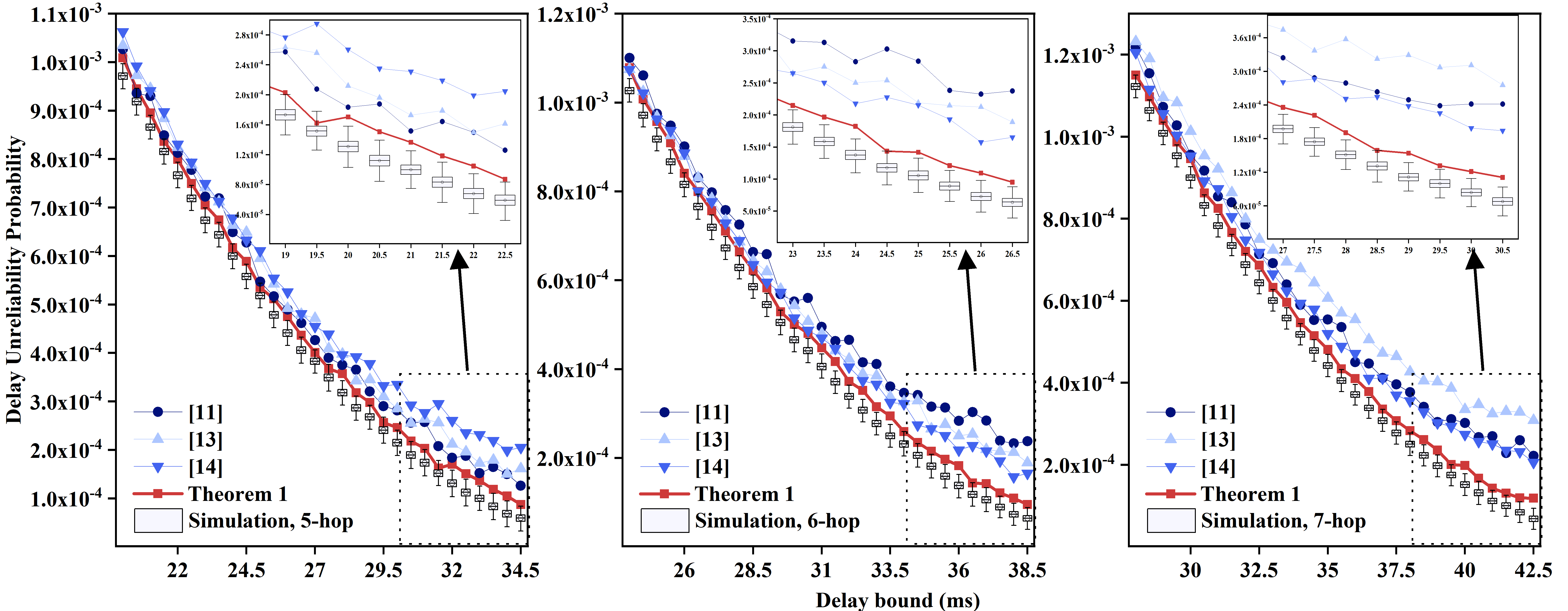}
\caption{Comparison of Delay unreliability probability bound compactness for Theorem \ref{theorem1}, \cite{rr24}, \cite{rr7}, \cite{rr6}. In multi-hop scenarios of 5 to 7 hops.}
\label{fig_3}
\end{figure}

Fig.~\ref{fig_2} illustrates the empirical delay unreliability probabilities (the box-chart labeled by 'Simulation') and theoretical bounds for $2$-$4$ hops. The proposed DUPB closely tracks simulation results, with a maximum deviation of $4.116\times{10}^{-5}$ for 4-hop at $W^b=22.5ms$. In contrast, the maximum deviation of each of these methods \cite{rr24,rr7,rr6} exceeds $9.4\times{10}^{-5}$. This validates the efficacy of the sliding block martingale in capturing fine-grained statistical features. As illustrated in the local zoomed-in views of each subfigure in Fig.~\ref{fig_2}, the proposed theoretical boundary consistently align closely with the Monte Carlo simulation results and exhibit better tightness compared to the methods in \cite{rr24,rr7,rr6}. To validate the scalability of the proposed theoretical boundaries in large-scale multi-hop scenarios, as shown in Fig.~\ref{fig_3}, with an increasing number of hops, these methods proposed in \cite{rr24,rr7,rr6} exhibits tail divergence, but the DUPB in Theorem \ref{theorem1} maintains consistency.

Moreover, we also calculated the Root Mean Square Error (RMSE) between the theoretical boundaries and the event occurrence probabilities, as shown in the Table.~\ref{tab:table1}. The RMSE of the theoretical boundaries derived from the proposed theorem is the smallest across all hop counts. In summary, the theoretical boundaries derived from the Theorem \ref{theorem1} exhibit good tightness and scalability. 

\begin{table}[h]
\footnotesize
\caption{Comparing the RMSE of different methods, RMSE describes the difference between the Delay Unreliability Probability and the DUPB.\label{tab:table1}}
\centering
\begin{tabular}{ccccccc}% 
\hline
 & \thead{2-hop\\$\left(\times{10}^{-5}\right)$} & \thead{3-hop\\$\left(\times{10}^{-5}\right)$} & \thead{4-hop\\$\left(\times{10}^{-5}\right)$} & \thead{5-hop\\$\left(\times{10}^{-5}\right)$} & \thead{6-hop\\$\left(\times{10}^{-5}\right)$} & \thead{7-hop\\$\left(\times{10}^{-5}\right)$}  \\
\hline
Theorem \ref{theorem1} & ${2.449}$	& 2.453	& 3.104	& 3.032 &	3.802 & 3.638\\
\cite{rr24} & 5.870 & 6.660 & 6.854 & 6.330 & 11.990 &	11.550\\
\cite{rr7} & 6.394	& 7.016	& 6.754	& 7.432	& 8.746	&	16.900\\
\cite{rr6} & 7.400	& 8.193	& 8.931	& 9.672	& 7.411	&	9.923\\
\hline
\end{tabular}
\end{table}

To evaluate the efficiency of the delay QoS analysis enhanced by the Lemma \ref{lemma1}, we conducted the following analysis. The components contributing to the delay QoS analysis time include: data collection and data analysis and computation. We primarily focus on the time consumed by data analysis and computation. In the data analysis and computation process, it is evident that the only iterative part involved in delay QoS analysis is the solution of $\theta$. 

Based on CPU time statistics, we found that in a single round of delay QoS analysis, the time spent on solving $\theta$ accounts for $70\%-90\%$ of the total analysis time, whether using the method from \cite{rr24,rr7,rr6} or the proposed method in this paper. Therefore, we focus on analyzing the iterative computation time for solving $\theta$. We used the fslove function provided by MATLAB to solve the $\theta$ problem given by (\ref{eq_dbmqos_13_1}) and compared it with the methods from \cite{rr24,rr7,rr6}.

Table \ref{tab:table2} compares the iterations required to solve $\theta$. The proposed (\ref{eq_dbmqos_13_1}) the average iteration count by $59\%$, $72\%$ and $58\%$ for 7-hop scenarios compared to \cite{rr24,rr7,rr6}, respectively. This efficiency stems from optimizing the iteration through Corollary \ref{corollary4}. Additionally, we provided the average CPU time statistics for each method in the Table \ref{tab:table3}, the proposed method reduces the average time consumption by $60.6\%$, $70.5\%$, and $59.8\%$, respectively, compared to the methods in \cite{rr24,rr7,rr6}. Based on the above, in $2$-$7$ hops scenarios, the iterative steps for solving $\theta$ using the Lemma \ref{lemma1} were consistently the fewest, and the CPU time consumed was the lowest.

\begin{table}[h]
\footnotesize
\caption{Comparison of computational efficiency of $\theta$ in different formulations, in the term of average iterations\label{tab:table2}}
\centering
\begin{tabular}{ccccccc}% 
\hline
 & \thead{2-hop,\\iterations} & \thead{3-hop,\\iterations} & \thead{4-hop,\\iterations} & \thead{5-hop,\\iterations} & \thead{6-hop,\\iterations} & \thead{7-hop,\\iterations} \\
\hline
(\ref{eq_dbmqos_13_1}) 	& $34$ & $43$ & $50$ & $38$  & $45$  & $47$\\

\cite{rr24} 					& $86$ & $109$ & $110$ & $94$  & $98$  & $115$\\

\cite{rr7} 						& $95$ & $121$ & $95$ & $153$  & $128$  & $170$\\

\cite{rr6} 						& $72$ & $96$ & $107$ & $103$  & $110$  & $112$\\
\hline
\end{tabular}
\end{table}

\begin{table}[h]
\footnotesize
\caption{Comparison of computational efficiency of $\theta$ in different formulations, in the term of average time consumption\label{tab:table3}}
\centering
\begin{tabular}{ccccccc}% 
\hline
 & 2-hop & 3-hop & 4-hop & 5-hop & 6-hop & 7-hop  \\
\hline
(\ref{eq_dbmqos_13_1}) & 78.78 &	53.09&	83.49&	77.94&	125.29	&65.98\\
\cite{rr24} & 188.38&	191.79&	209.89&	194.94&	236.89&	209.28\\
\cite{rr7} & 207.38&	216.89&	178.29&	320.34&	299.49&	324.08\\
\cite{rr6} & 158.88&	164.89&	203.19&	214.44&	262.09&	202.88\\
\hline
\end{tabular}
\end{table}

For the effectiveness of the minimum service rate given in (\ref{eq_4_18a}), we recorded the minimum service rate and the average downlink data rate (i.e., average rate of traffic from Marco Base Station.) of the UE under different service requirements. The results in Table~\ref{tab:table4} confirm that the minimum service rate in (\ref{eq_4_18a}) effectively supports services with varying strengths and QoS guarantee requirements in multi-hop transmission with different hops. Moreover, when the intensity of the traffic is close, the smaller the delay bound, the larger the difference between the minimum service rate and the average downlink data rate. This indicates that the results derived from (\ref{eq_4_18a}) can accommodate different levels of QoS guarantee and have the capability to provide guidance for bandwidth allocation.

\begin{table}[t]
\caption{minimum service rate and average downlink data rate for different service requirements\label{tab:table4}}
\centering
\begin{tabular}{ccc}
\hline
\thead{Service type \\and QoS requirement}& \thead{Minimum service\\rate}& \thead{average downlink\\data rate}\\
\hline
\thead{16Mbps,$W^b=10{\rm ms}$,$\varepsilon=1e^{-5}$}&
$8281.2{\rm bits/slot}$ &$8001.7{\rm bits/slot}$\\
\thead{16Mbps,$W^b=20{\rm ms}$,$\varepsilon=1e^{-5}$}&
$7994.3{\rm bits/slot}$ &$7825.3{\rm bits/slot}$\\

\thead{11Mbps,$W^b=10{\rm ms}$,$\varepsilon=1e^{-5}$}&
$5694.3{\rm bits/slot}$ &$5329.7{\rm bits/slot}$\\
\thead{11Mbps,$W^b=20{\rm ms}$,$\varepsilon=1e^{-5}$}&
$5652.5{\rm bits/slot}$ &$5525.3{\rm bits/slot}$\\

\thead{7Mbps,$W^b=10{\rm ms}$,$\varepsilon=1e^{-5}$}&
$3785.7{\rm bits/slot}$ &$3583.1{\rm bits/slot}$\\
\thead{7Mbps,$W^b=20{\rm ms}$,$\varepsilon=1e^{-5}$}&
$3511.4{\rm bits/slot}$ &$3463.2{\rm bits/slot}$\\
\hline
\end{tabular}
\end{table}

The simulation results validate three key aspects of the proposed framework. First, the sliding block martingale-based DUPB (Theorem \ref{theorem1}) demonstrates tight theoretical bounds across multi-hop scenarios, the RMSE metrics further support this. Second, the $\theta$-solving method significantly improves computational efficiency, reducing iterations and CPU time, enhancing efficiency of QoS analysis. Lastly, The derived minimum service rates are well suited to support traffic with different mean rates and QoS guarantees. These results highlight the good accuracy, analysis efficiency and scalability of this QoS analysis method in multi-hop networks.

\section{Conclusion}
This paper proposes a multi-hop delay QoS analysis framework based on sliding block martingale, which improves the accuracy and efficiency of delay QoS analysis in multi-hop networks. The constructed sliding block martingale captures both short-term fluctuations and long-term trends of the stochastic process through a sliding window, eliminating the reliance on generalized increasing property. Based on the cascade property of multi-hop queues and combining it with sliding block martingale, we derive a more compact DUPB (Theorem \ref{theorem1}). In scenarios with 2 to 4 hops, the maximum deviation between the theoretical bound and Monte Carlo simulation results is $4.116 \times 10^{-5}$, which is better than other methods. When extended to 7 hops, the theoretical bound remains compact without tail divergence. In the 2 to 7 hops scenarios, the maximum RMSE of the theoretical bound is $3.802 \times 10^{-5}$ (6-hop), which is lower than that of other methods. To address the issue that the original queue steady-state condition become unavailable due to changes in the martingale construction, we propose Lemma \ref{lemma1} to reconstruct the queue steady-state condition, and introduce the Micrometric Intervals-based Supermartingale Upcrossing Estimate Theorem (Theorem \ref{theorem2}) along with its corollary (Corollary \ref{corollary4}), which quantifies the upper bound on the event occurrence frequency and narrows the solution space of the key parameter $\theta$. Simulation results show that the number of iterations for solving $\theta$ and the CPU computation time for 2 to 7 hops are lower than other methods, enhancing QoS analysis efficiency. Furthermore, based on Theorem \ref{theorem1} and the cascade property of multi-hop queues, we derive the explicit expression for the minimum service rate of multi-hop queues, which shows that the minimum service rate is always greater than the average arrival rate of users under different traffic intensities and QoS guarantee requirements, providing an effective reference for bandwidth and power adjustment. In summary, the sliding block martingale based multi-hop delay QoS analysis framework proposed in this paper demonstrates excellent compactness, scalability, and computational efficiency in multi-hop delay QoS analysis.

\appendix
\section{Proof of correctness of constructing Martingales based on 
backlog}
\label{apped:pfcmb}
% 在证明开始之前, 首先要说明假设的是瞬时到达随机过程和瞬时服务随机过程是一个独立同分布的随机过程
% 为什么要证明随机过程是严平稳过程
% 构造积压鞅的前提是随机变量序列是一个非负且相互独立的, 以及随机过程是一个严平稳随机过程
Constructing the backlog martingales requires that the r.v. $\left\{{rv}_{q(t)}\mid t\in\mathbb{N}^{+}\right\}$ is non-negative and independent, and that $\left\{q\left(t\right),t\in\mathbb{N}^{+}\right\}$ must be a stationary random process.
% 先写bt的表达式
Firstly, derive the formulation of $q\left(t\right)$. In \cite{r39}, the Lindley equation is used to derive $Q\left(t\right)$, which is given by
\begin{equation}
\label{eq_pfcmbob_1}
\begin{array}{lll}
Q\left(t\right)=\max\left\{0,Q\left(t-1\right)+a\left(t\right)-s\left(t\right)\right\}.
\end{array}
\end{equation}

Then, the instantaneous backlog rate $q\left(t\right)$ is calculated as
\begin{equation}
\label{eq_pfcmbob_2}
\begin{array}{lll}
q\left(t\right)
=Q\left(t\right)-Q\left(t-1\right)\\
=\max\left\{0,Q\left(t-1\right)+a\left(t\right)-s\left(t\right)\right\}-Q\left(t-1\right)\\
\overset{a}{=}\left(Q\left(t-1\right)+a\left(t\right)-s\left(t\right)\right)\mathbf{1}_{\left\{Q\left(t-1\right)+a\left(t\right)-s\left(t\right)> 0\right\}}-Q\left(t-1\right)\\
=\left(Q\left(t-1\right)+a\left(t\right)-s\left(t\right)\right)\mathbf{1}_{\left\{Q\left(t-1\right)+a\left(t\right)-s\left(t\right)> 0\right\}}\\
\ \ \ -Q\left(t-1\right)\left(\mathbf{1}_{\left\{Q\left(t-1\right)+a\left(t\right)-s\left(t\right)> 0\right\}}+\mathbf{1}_{\left\{Q\left(t-1\right)+a\left(t\right)-s\left(t\right)\le 0\right\}}\right)\\
=\left(a\left(t\right)-s\left(t\right)\right)\mathbf{1}_{\left\{Q\left(t-1\right)+a\left(t\right)-s\left(t\right)> 0\right\}}\\
\ \ \ -Q\left(t-1\right)\mathbf{1}_{\left\{Q\left(t-1\right)+a\left(t\right)-s\left(t\right)\le 0\right\}}\\
\overset{b}{=}\left(a\left(t\right)-s\left(t\right)\right)\mathbf{1}_{\left\{Q\left(t\right)> 0\right\}}-Q\left(t-1\right)\mathbf{1}_{\left\{Q\left(t\right)= 0\right\}}.
\end{array}
\end{equation}
% 指示函数将随机变量的可测空间限定在$$内, 而当$$时, $=0$, 所以step holds.
$\mathbf{1}\left(\cdot \right)$ restricts the measurable space of the r.v. $Q\left(t-1\right)+a\left(t\right)-s\left(t\right)$ to $\left\{Q\left(t-1\right)+a\left(t\right)-s\left(t\right)> 0\right\}$. Consequently $\mathbf{1}\left(Q\left(t-1\right)+a\left(t\right)-s\left(t\right)> 0\right)$ is equal to 0 when $Q\left(t-1\right)+a\left(t\right)-s\left(t\right)\le 0$, thus $a$ in (\ref{eq_pfcmbob_2}) holds.
From (\ref{eq_pfcmbob_1}), $Q\left(t-1\right)+a\left(t\right)-s\left(t\right)> 0$ and $Q\left(t\right)>0$ are equivalent, and $Q\left(t-1\right)+a\left(t\right)-s\left(t\right)\le 0$ and $Q\left(t\right)=0$ are equivalent, so $b$ in (\ref{eq_pfcmbob_2}) holds.
According to (\ref{eq_pfcmbob_2}), when the backlog is non-zero, $q\left(t\right)=a\left(t\right)-s\left(t\right)$. % Then, the following analyses are conducted under the assumption that the backlog is non-zero. 

% The base of backlog-martingales construction is the independent and non-negative properties.
Secondly, prove that $\left\{q\left(t\right)\mid t\in\mathbb{N}^{+}\right\}$ in  is independent when backlog is non-empty. For $\forall t,i\in \mathbb{N}^{+}$, the covariance between $q\left(t\right)$ and $q\left(t+i\right)$ is
\begin{equation}
\label{eq_pfcmbob_3}
\begin{array}{l}
{\rm Cov}\left(q\left(t\right),q\left(t+i\right)\right)
=\mathbb{E}\left[q\left(t\right)q\left(t+i\right)\right]-\mathbb{E}\left[q\left(t\right)\right]\mathbb{E}\left[q\left(t+i\right)\right]\\
% =\mathbb{E}\left[\left(a\left(t\right)-s\left(t\right)\right)\left(a\left(t+i\right)-s\left(t+i\right)\right)\right]-\mathbb{E}\left[b\left(t\right)\right]\mathbb{E}\left[b\left(t+i\right)\right]\\
=\mathbb{E}\left[\left(a\left(t\right)-s\left(t\right)\right)\left(a\left(t+i\right)-s\left(t+i\right)\right)\right]\\
\ \ \ -\mathbb{E}\left[a\left(t\right)-s\left(t\right)\right]\mathbb{E}\left[a\left(t+i\right)-s\left(t+i\right)\right]\\
=\mathbb{E}\left[a\left(t\right)a\left(t+i\right)+s\left(t\right)s\left(t+i\right)-s\left(t\right)a\left(t+i\right)-s\left(t+i\right)a\left(t\right)\right]\\
\ \ \ -\mathbb{E}\left[a\left(t\right)\right]\mathbb{E}\left[a\left(t+i\right)\right]-\mathbb{E}\left[s\left(t\right)\right]\mathbb{E}\left[s\left(t+i\right)\right]\\
\ \ \ +\mathbb{E}\left[s\left(t\right)\right]\mathbb{E}\left[a\left(t+i\right)\right]+\mathbb{E}\left[a\left(t\right)\right]\mathbb{E}\left[s\left(t+i\right)\right]\\
\overset{a}{=}\mathbb{E}\left[a\left(t\right)a\left(t+i\right)+s\left(t\right)s\left(t+i\right)-s\left(t\right)a\left(t+i\right)-s\left(t+i\right)a\left(t\right)\right]\\
\ \ \ -\mathbb{E}\left[a\left(t\right)a\left(t+i\right)\right]-\mathbb{E}\left[s\left(t\right)s\left(t+i\right)\right]\\
\ \ \ +\mathbb{E}\left[s\left(t\right)a\left(t+i\right)\right]+\mathbb{E}\left[a\left(t\right)s\left(t+i\right)\right]=0.
\end{array}
\end{equation}
Both instantaneous arrival $\left\{a\left(t\right)\mid t\in\mathbb{N}^{+}\right\}$ and instantaneous service $\left\{s\left(t\right)\mid t\in\mathbb{N}^{+}\right\}$ are i.i.d., so $a$ in (\ref{eq_pfcmbob_3}) holds. Furthermore the arrival is independent of the service at any given time, thus $\left\{q\left(t\right)\mid t\in\mathbb{N}^{+}\right\}$ is independent. Consequently $\left\{{rv}_{q(t)}\mid i\in\mathbb{N}^{+}\right\}$ is independent and non-negative.% exponential function is non-negative

% 求解随机序列的期望\方差\协方差, 确定随机序列是时不变的, 证明随机序列是宽平稳的
Thirdly, prove that $\left\{q\left(t\right)\mid t\in\mathbb{N}^{+}\right\}$ is a stationary.
Based on (\ref{eq_pfcmbob_2}), the expectation and variance of $\left\{q\left(t\right)\mid t\in\mathbb{N}^{+}\right\}$ are as follows
\begin{equation}
\label{eq_pfcmbob_4}
\begin{array}{l}
\mathbb{E}\left[q\left(t\right)\right]=\mathbb{E}\left[a\left(t\right)\right]-\mathbb{E}\left[s\left(t\right)\right]\\
{\rm Var}\left[q\left(t\right)\right]=\mathbb{E}\left[q^2\left(t\right)\right]-\left(\mathbb{E}\left[b\left(t\right)\right]\right)^2\\
=\mathbb{E}\left[a^2\left(t\right)+s^2\left(t\right)\right]-\left(\mathbb{E}\left[a\left(t\right)\right]\right)^2-\left(\mathbb{E}\left[s\left(t\right)\right]\right)^2\\
={\rm Var}\left[a\left(t\right)\right]+{\rm Var}\left[s\left(t\right)\right].
\end{array}
\end{equation}
Obviously, both $\mathbb{E}\left[q\left(t\right)\right]$ and ${\rm Var}\left[q\left(t\right)\right]$ are non-time-varying. Then, take two sequences with the length $L\in\mathbb{N}^{+}$ from $\left\{q\left(t\right)\mid t\in\mathbb{N}^{+}\right\}$, ${seq}_1=\left[q\left(t_1\right),\ldots,q\left(t_L\right)\right]$ and ${seq}_2=\left[q\left(t_1+h\right),\ldots,q\left(t_L+h\right)\right]$, the covariance matrix of ${seq}_1$ and ${seq}_2$ are denoted as $\mathbf{Cov}\left({seq}_1\right)\in\mathbb{R}^{L\times L}$ and $\mathbf{Cov}\left({seq}_2\right)\in\mathbb{R}^{L\times L}$, respectively. The elements in these two matrices are $\mathbf{Cov}\left({seq}_1\right)_{i,j}={\rm Cov}(q\left(t_i\right),q\left(t_j\right))$ and $\mathbf{Cov}\left({seq}_2\right)_{i,j}={\rm Cov}(q\left(t_i+h\right),q\left(t_j+h\right))$.
% \begin{equation}
% \label{eq_pfcmbob_5}
% \begin{array}{l}
% \mathbf{Cov}\left({seq}_1\right)_{i,j}={\rm Cov}(b\left(t_i\right),b\left(t_j\right))\\
% \mathbf{Cov}\left({seq}_2\right)_{i,j}={\rm Cov}(b\left(t_i+h\right),b\left(t_j+h\right))
% \end{array}
% \end{equation}
The expectation vector of ${seq}_1$ and ${seq}_2$ are $\mathbb{E}\left[{seq}_1\right]=\left[\mathbb{E}\left[q\left(t_1\right)\right],\ldots,\mathbb{E}\left[q\left(t_L\right)\right]\right]$ and $\mathbb{E}\left[{seq}_2\right]=\left[\mathbb{E}\left[q\left(t_1+h\right)\right],\ldots,\mathbb{E}\left[q\left(t_L+h\right)\right]\right]$, respectively.
% \begin{equation}

% \label{eq_pfcmbob_5}
% \begin{array}{l}
% \mathbb{E}\left[{seq}_1\right]=\left[\mathbb{E}\left[b\left(t_1\right)\right],\ldots,\mathbb{E}\left[b\left(t_L\right)\right]\right]\\
% \mathbb{E}\left[{seq}_2\right]=\left[\mathbb{E}\left[b\left(t_1+h\right)\right],\ldots,\mathbb{E}\left[b\left(t_L+h\right)\right]\right]\\
% \end{array}
% \end{equation}
From (\ref{eq_pfcmbob_3}), for $\forall t,\forall i\in \mathbb{N}^{+}$, the covariance between $q\left(t\right)$ and $q\left(t+i\right)$ is equal to 0, such that, ${seq}_1$ and ${seq}_2$ have the same covariance matrix. Additionally, based on (\ref{eq_pfcmbob_4}), and the fact that $\left\{a\left(t\right)\mid t\in\mathbb{N}^{+}\right\}$ and $\left\{s\left(t\right)\mid t\in\mathbb{N}^{+}\right\}$ are i.i.d., ${seq}_1$ and ${seq}_2$ have the same expectation vector. Thus, the random process $\left\{q\left(t\right)\mid t\in\mathbb{N}^{+}\right\}$ is a stationary random process.
Finally, for $D_Q\left(\theta\right)$, let's assumed that $\mathbb{E}\left[{rv}_{q(t)}\right]=\mathbb{E}\left[{e^{\theta\left(q\left(t\right)-D_Q\left(\theta\right)\right)}}\right]=1$, and then
\begin{equation}
\label{eq_pcbm_3}
\begin{array}{lll}
% &\mathbb{E}\left[e^{\theta\left[b\left(t\right)-K_B\left(\theta\right)\right]}\right]=1\\
% \Leftrightarrow &\mathbb{E}\left[e^{\theta b\left(t\right)}\right]=e^{\theta K_B\left(\theta\right)}\\
% \Leftrightarrow 
\ln\left(\mathbb{E}\begin{bmatrix}e^{\theta q\left(t\right)}\end{bmatrix}\right)=\theta D_Q\left(\theta\right)
\Leftrightarrow D_Q\left(\theta\right)=\frac{1}{\theta}\ln\left(\mathbb{E}\begin{bmatrix}e^{\theta q\left(t\right)}\end{bmatrix}\right)
\end{array}
\end{equation}
The random process consist of the element that the products of non-negative independent r.v. of mean $1$, this random process is the martingale \cite{r1}.

\section{Proof of correctness of constructing Sliding Block Martingale}
\label{apped:pfcdm}
The deterministic function is $D_{X}\left(\theta\right)=\frac{1}{\theta
 }\ln\mathbb{E}\begin{bmatrix}e^{\theta\left(\frac{X\left(t+{W}^b\right)-X\left(t\right)}{{W}^b}\right)}\end{bmatrix}$. 
The expectation of ${rv}_{x(t)}$ 
\begin{equation}
\label{eq_pcdbm_2}
\begin{array}{lll}
\mathbb{E}\left[e^{\theta\left(x\left(t\right)-D_{X}\left(\theta\right)\right)}\right]=\mathbb{E}\begin{bmatrix}
e^{\left({\theta\left[x\left(t\right)-{\frac{1}{\theta
 }\ln\mathbb{E}\begin{bmatrix}e^{\theta\left(\frac{X\left(t+{W}^b\right)-X\left(t\right)}{{W}^b}\right)}\end{bmatrix}}\right]}\right)}
\end{bmatrix}\\
\overset{a}{=}\mathbb{E}\begin{bmatrix}e^{\theta x\left(t\right)}\end{bmatrix}\cdot \mathbb{E}\begin{bmatrix}
e^{-\ln\mathbb{E}\begin{bmatrix}e^{\theta\left(\frac{X\left(t+{W}^b\right)-X\left(t\right)}{{W}^b}\right)}\end{bmatrix}} 
\end{bmatrix}\\
=\mathbb{E}\begin{bmatrix}e^{\theta x\left(t\right)}\end{bmatrix}\cdot \mathbb{E}\begin{bmatrix}
\left(\mathbb{E}\begin{bmatrix}e^{\theta\left(\frac{X\left(t+{W}^b\right)-X\left(t\right)}{{W}^b}\right)}\end{bmatrix}\right)^{-1} 
\end{bmatrix}\\
\overset{b}{=}\mathbb{E}\begin{bmatrix}e^{\theta x\left(t\right)}\end{bmatrix}\cdot \left(\mathbb{E}\begin{bmatrix}e^{\theta\left(\frac{X\left(t+{W}^b\right)-X\left(t\right)}{{W}^b}\right)}\end{bmatrix}\right)^{-1}
\overset{c}{=}1.
\end{array}
\end{equation}
$\ln\mathbb{E}\begin{bmatrix}e^{\theta\left(\frac{X\left(t+{W}^b\right)-X\left(t\right)}{{W}^b}\right)}\end{bmatrix}$ is constant, so $a$ holds. % Obviously, $\mathbb{E}[g(\mathbb{E}[x])]=g(\mathbb{E}[x])$, 
The expectation of a constant is the constant itself, i.e., $\mathbb{E}[g(\mathbb{E}[x])]=g(\mathbb{E}[x])$, so $b$ holds.
For step $c$ in (\ref{eq_pcdbm_2}), we provide the following explanation:
$\frac{X\left(t+{W}^b\right)-X\left(t\right)}{{W}^b}$ denotes the average of $\left\{x\left(\hat{t}\right)\mid \hat{t}\in\mathbb{N}^{+}\right\}$ over the time interval $\left[t,t+{W}^b\right]$. This can be expressed as $\mathbb{E}\begin{bmatrix}x\left(t\right);F_{t+{W}^b}\end{bmatrix}$, where the $F_{t+{W}^b}$ denotes the event that $x\left(t\right)$ is constrained to $\left[t,t+W^b\right]$.
In fact, $\mathbb{E}\begin{bmatrix}x\left(t\right);F_{t+{W}^b}\end{bmatrix}$ is a conditional expectation and can also be written as $\mathbb{E}\begin{bmatrix}x\left(t\right)|\sigma(F_{t+{W}^b})\end{bmatrix}$, where $\sigma(F_{t+{W}^b})$ is the sigma-algebra generated by $F_{t+{W}^b}$. Therefore, based on the property of conditional expectation, we have
\begin{equation}
\label{eq_pcdbm_3}
\begin{array}{lll}
\mathbb{E}\begin{bmatrix}e^{\theta\left(\frac{X\left(t+{W}^b\right)-W\left(t\right)}{{W}^b}\right)}\end{bmatrix}
=\mathbb{E}\begin{bmatrix}e^{\theta \mathbb{E}\begin{bmatrix}x\left(\hat{t}\right);F_{t+{W}^b}\end{bmatrix}}\end{bmatrix}\\
=\mathbb{E}\begin{bmatrix}e^{\theta \mathbb{E}\begin{bmatrix}x\left(\hat{t}\right)|\sigma(F_{t+{W}^b})\end{bmatrix}}\end{bmatrix}
\overset{a}{=}\mathbb{E}\begin{bmatrix}e^{\theta x\left(\hat{t}\right)}\end{bmatrix}.
\end{array}
\end{equation}
Since $\left\{x\left(\hat{t}\right)\mid \hat{t}\in\left[t,t+{W}^b\right]\right\}$ is $\sigma(F_{t+{W}^b})$-measurable, such that $\mathbb{E}\begin{bmatrix}x\left(\hat{t}\right)|\sigma(F_{t+{W}^b})\end{bmatrix}=x\left(\hat{t}\right)$, so step $a$ in (\ref{eq_pcdbm_3}) holds. Moreover, $\left\{x\left(\hat{t}\right)\mid \hat{t}\in\mathbb{N}^{+}\right\}$ is a stationary random process, i.e., $\mathbb{E}\begin{bmatrix}e^{\theta x\left(\hat{t}\right)}\end{bmatrix}=\mathbb{E}\begin{bmatrix}e^{\theta x\left({t}\right)}\end{bmatrix}$, so step $c$ in (\ref{eq_pcdbm_2}) holds.
As discussed at the end of \ref{apped:pfcmb}, it is evident that $ {M_{SB}}\left(X\left(t\right),t,\theta,{W}^b\right)=\prod\limits_{i=t}^{t+{W}^b}{e^{\theta (x\left(i\right)-D_{X}\left(\theta\right))}}$ is martingale.

\section{Proof of Theorem \ref{theorem1}}
\label{apped:pt2}
Based on (\ref{eq_dbmqos_11_b2}) and (\ref{eq_dbmqos_11_b3}), the step $a$ in (\ref{eq_pmdupb_3}) is holds. The step ${b}$ in (\ref{eq_pmdupb_3}) is derived based on the Markov's inequality. 
\begin{equation}
\label{eq_pmdupb_3}
\begin{array}{llll}
&\mathbb{P}\left\{W\left(t\right)\ge {W}^b\right\} \le \mathbb{P}\left\{
A_1\left(t\right) 
- {A_{{hop}}^*}\left(t+{W}^b\right)\ge 0\right\}\\
\overset{a}{=}&\mathbb{P}\left\{
\begin{array}{ccc}
\sum\limits_{i=1}^{{hop}}Q_{i}\left(t+{W}^b \right)\ge 
A_1\left(t+{W}^b\right)-A_1\left(t\right)
\end{array}
\right\}\\
=& \mathbb{P}\left\{
\begin{array}{ccc}
\sum\limits_{i=1}^{{hop}}\left(Q_{i}\left(t+{W}^b \right)-Q_{i}\left(t \right)\right)\ge \\
A_1\left(t+{W}^b\right)-A_1\left(t\right)-\sum\limits_{i=1}^{{hop}}Q_{i}\left(t \right)
\end{array}
\right\}\\
=& \mathbb{P}\left\{
\begin{array}{ccc}
\sum\limits_{i=1}^{{hop}}\left(Q_{i}\left(t+{W}^b \right)-Q_{i}\left(t \right)-{W}^b{D_{Q_{i}}\left(\theta\right)}\right)\\
-\left(A_1\left(t+{W}^b\right)-A_1\left(t\right)-{{W}^b}D_{A_1}\left(\theta\right)\right)\\
\ge{{W}^b}D_{A_1}\left(\theta\right)-\sum\limits_{i=1}^{{hop}}\left(Q_{i}\left(t \right)+{{W}^b}{D_{Q_{i}}\left(\theta\right)}\right)\\
\end{array}
\right\}\\
{=}& \mathbb{P}\left\{
\begin{array}{ccc}
\frac{\prod\limits_{i=1}^{{hop}}{{M_{SB}}\left(Q_{i}\left(t\right),t,\theta,{W^b}\right)}}{{M_{SB}}\left(A_{1}\left(t\right),t,\theta,{{W}^b}\right)}\ge 

\frac{e^{\theta {{W}^b} D_{A_1}\left(\theta\right)}}
{e^{ 
\theta \sum\limits_{i=1}^{{hop}}{\left(Q_{i}\left(t \right)+{{W}^b}{D_{B_i}\left(\theta\right)}\right)}
}}
\end{array}
\right\}\\
\overset{b}{\le}&\begin{array}{ccc}\mathbb{E}\begin{bmatrix}\frac{\prod\limits_{i=1}^{{hop}}{{M_{SB}}\left(Q_{i}\left(t\right),t,\theta,{W^b}\right)}}{{M_{SB}}\left(A_{1}\left(t\right),t,\theta,{{W}^b}\right)}\end{bmatrix}\frac{e^{ \theta \sum\limits_{i=1}^{{hop}}{\left(Q_{i}\left(t \right)+{{W}^b}{D_{B_i}\left(\theta\right)}\right)}}}{e^{\theta {{W}^b} D_{A_1}\left(\theta\right)}}
\end{array}.
\end{array}
\end{equation}

Since the backlog at multi-hop are not independent of each other, then
\begin{equation}
\label{eq_pmdupb_4}
\begin{array}{llll}
\mathbb{E}\begin{bmatrix}\frac{\prod\limits_{i=1}^{{hop}}{{M_{SB}}\left(Q_{i}\left(t\right),t,\theta,{W^b}\right)}}{{M_{SB}}\left(A_{1}\left(t\right),t,\theta,{{W}^b}\right)}\end{bmatrix}
=\mathbb{E}\begin{bmatrix}\frac{{e^{\theta\left(\sum\limits_{i=1}^{{hop}}\left(Q_i\left(t+{W}^b\right)-Q_i\left(t\right)\right)-A_1\left(t+{W}^b\right)+A_1\left(t\right)\right)}}}{e^{\theta {W}^b\left(\sum\limits_{i=1}^{{hop}}D_{B_i}\left(\theta\right)-D_{A_1}\left(\theta\right)\right)}}\end{bmatrix}\\
=e^{\theta {W}^b\left(D_{A_1}\left(\theta\right)-\sum\limits_{i=1}^{{hop}}D_{B_i}\left(\theta\right) \right)}
\mathbb{E}\begin{bmatrix}
{e^{\theta X_{msb}}}
\end{bmatrix}.\\
\end{array}
\end{equation}
where the $X_{msb}:=\sum\limits_{i=1}^{{hop}}\left(Q_i\left(t+{W}^b\right)-Q_i\left(t\right)\right)-A_1\left(t+{W}^b\right)+A_1\left(t\right)$, and $\mathbb{E}\left[{e^{\theta X_{msb}}}\right]$ is derived from long-term statistics of backlogs and arrivals. Finally, we have:
\begin{equation}
\label{eq_pmdupb_5}
\begin{array}{llll}
\mathbb{P}\left\{
W\left(t\right)\ge {W}^b
\right\}\le
\mathbb{E}\begin{bmatrix}{e^{\theta X_{msb}}}\end{bmatrix}e^{ \theta \sum\limits_{i=1}^{{hop}}{Q_{i}\left(t\right)}}.
\end{array}
\end{equation}

\section{Proof of the Lemma \ref{lemma1}}
\label{apped:ssc}
The entire multi-hop queue can be viewed as a complete single-hop queue based on the concatenation property, the queue's steady-state condition ($\mathbb{E}\left[s(t)\right]\ge\mathbb{E}\left[a(t)\right]$) is equivalent to $S(\hat{t})\ge A(\hat{t})$, then
\begin{equation}
\label{eq_psteady_1}
\begin{array}{lll}
S(\hat{t})\ge A(\hat{t})\Leftrightarrow 
A\left(\hat{t}+{W}^b\right)-S(\hat{t})\le A\left(\hat{t}+{W}^b\right)-A(\hat{t}),\\
\end{array}
\end{equation}
and based on the definition $A^*\left(\hat{t}+{W}^b\right)$ \cite{r39}, 
\begin{equation}
\label{eq_psteady_2}
\begin{array}{lll}
A^*\left(\hat{t}+{W}^b\right)=\underset{0\le s\le \hat{t}+{W}^b}{\inf}\left\{A(s)+S(s,\hat{t}+{W}^b)\right\}\\
\overset{a}{\ge} \underset{0\le s\le \hat{t}+{W}^b}{\inf}\left\{S(s)+S(s,\hat{t}+W^b)\right\}= S(\hat{t}+{W}^b)\ge S(\hat{t})\\ \Leftrightarrow
Q\left(\hat{t}+{W}^b\right)=A\left(\hat{t}+{W}^b\right)-A^*\left(\hat{t}+{W}^b\right)\le A\left(\hat{t}+{W}^b\right)-S(\hat{t})\\
\end{array}
\end{equation}
The cumulative service in the queue at any time will not exceed the cumulative arrivals (i.e., $A(s)\ge S(s)$), so step $a$ in (\ref{eq_psteady_2}) holds. Please note that the steady-state condition serves as a criterion for solving $\theta$ rather than reflecting the actual physical state of the queue in the real world. Therefore, the inequality $A(s)\ge S(s)$ does not contradict the equivalent steady state condition $S(\hat{t})\ge A(\hat{t})$. Thus, the steady state condition can be transformed as $Q\left(\hat{t}+{W}^b\right)\le A\left(\hat{t}+{W}^b\right)-A(\hat{t})$. 

In the multi-hop transmission shown in Fig.\ref{fig_0}, $Q\left(\hat{t}+{W}^b\right)$ is in fact $\sum\limits_{i=1}^{{hop}}{Q_{i}\left(\hat{t}+{W}^b \right)}$, and the arrival of this entire multi-hop queue is $A_1$. Thus, the steady-state condition can be expressed as $\sum\limits_{i=1}^{{hop}}{Q_{i}\left(t+{W}^b \right)}\le A_1\left(t+{W}^b\right)-A_1\left(t\right)$.  
By the definition of sliding block martingale, $D_{Q_i}\left(\theta\right)$ and $D_{A_1}\left(\theta\right)$ represent the average rates of $Q_i\left(t\right)$ and $A_1\left(t\right)$ over $\left[t,t+{W}^b\right]$, respectively. Consequently, the steady-state condition becomes $\sum\limits_{i=1}^{{hop}}{{W}^b{D_{Q_i}\left(\theta\right)}}+\sum\limits_{i=1}^{{hop}}Q_{i}\left(t \right)\le {{W}^b}D_{A_1}\left(\theta\right)$. 

\section{Proof of Theorem \ref{theorem2}}
\label{apped:pt3}
Assumed that $Y_T$ is a supermartingale and bounded in $\mathcal{L}^1$, 
and $U_T\left({sb}_{l}^{i},{sb}_{h}^{i}\right)$ is the number of upcrossing $\left[{sb}_{l}^{i},{sb}_{h}^{i}\right]$, where
\begin{equation}
\label{eq_pmdu_1}
\begin{array}{llll}
U_T\left({sb}_{l}^{i},{sb}_{h}^{i}\right):=
\max\left\{
k\mid
\begin{array}{ll}
\exists 0\le s_1<t_1<\ldots<s_k<t_k\le N \\ % s_2<t_2<
Y_{s_i}<{sb}_{l}^{i},Y_{t_i}>{sb}_{h}^{i},\forall i\in\left[1,k\right]
\end{array}
\right\}.\\
\end{array}
\end{equation}
Based on the Doob's Upcrossing Lemma, we have $\mathbb{E}\left[U_T\left({sb}_{l}^{i},{sb}_{h}^{i}\right)\right]\le\frac{\mathbb{E}\left[\left(Y_t-{sb}_{l}^{i}\right)^{-}\right]}{{sb}_{h}^{i}-{sb}_{l}^{i}}$. 
To ensure that each upcrossing sub-interval captures the variation in $Y_T$, the length of each sub-interval is set to the expectation of the difference between adjacent points of $Y_T$, denoted as $\delta =\mathbb{E}\left[Y_t-Y_{t-1}\right]$. Consequently, the interval is divided as shown in (\ref{eq_dbmqos_13_b2}).
Analogous to integral, ${O_T}\left[a\right]$ is bounded by the sum of the upcrossing counts for each sub-interval, i.e., 
\begin{equation}
\label{eq_pmdu_4}
\begin{array}{lll}
{O_T}\left[a\right]
&\le\sum\limits_{i=1}^{{sb}_{seg}}{\frac{2\mathbb{E}\left[\left(Y_t-{sb}_{l}^{i}\right)^{-}\right]}{\delta }}.\\
\end{array}
\end{equation}
The factor of $2$ arises because, for a supermartingale, the conditional expectation gradual decreases, thus, a upcrossing in a subinterval indicates a future downcrossing within that subinterval.

\section{Proof of Corollary \ref{corollary4}}
\label{apped:c1}
For the delay unreliability necessary random process $Y_T^{delay}$, 
\begin{equation}
\label{eq_pumfdu_1}
\begin{array}{lll}
\mathbb{E}\left[Y_t^{delay}\mid\mathcal{F}_{t-1}\right]=
\mathbb{E}\left[Y_{t-1}^{delay}\mid\mathcal{F}_{t-1}\right]+\\
\mathbb{E}\left[\sum\limits_{i=1}^{{hop}}{q_i\left(t+{W}^b\right)}-a_1\left(t+{W}^b\right)+a_1\left(t\right)\right],
% \end{array}\\
\end{array}
\end{equation}
where the $\mathcal{F}_{t-1}$ is the filtration of $Y_T^{delay}$ up to state $t-1$.

\begin{equation}
\label{eq_pumfdu_1_1}
\begin{array}{lll}
\mathbb{E}\left[\sum\limits_{i=1}^{{hop}}{q_i\left(t+{W}^b\right)}-a_1\left(t+{W}^b\right)+a_1\left(t\right)\right]\\
\overset{a}{=}\mathbb{E}\left[\sum\limits_{i=1}^{{hop}}{(a_i\left(t+{W}^b\right)-s_i\left(t+{W}^b\right))}-a_1\left(t+{W}^b\right)+a_1\left(t\right)\right]\\
\overset{}{=}\mathbb{E}\left[\sum\limits_{i=1}^{{hop}}{(a_i\left(t+{W}^b\right)-s_i\left(t+{W}^b\right))}\right]-\mathbb{E}\left[a_1\left(t+{W}^b\right)\right]+\mathbb{E}\left[a_1\left(t\right)\right]\\
\overset{}{=}\sum\limits_{i=1}^{{hop}}(\mathbb{E}\left[a_i\left(t+{W}^b\right)\right]-\mathbb{E}\left[s_i\left(t+{W}^b\right)\right])\overset{b}{\le}0.
\end{array}
\end{equation}
% \begin{equation}
% \label{eq_pumfdu_2}
% \begin{array}{lll}
% \mathbb{E}\left[X_n^{delay}|\mathcal{F}_{n-1}\right]\le X_{n-1}^{delay}
% \end{array}
% \end{equation}
Then, as shown in \ref{apped:pfcmb}, when the backlog is non-zero, $q\left(t\right)=a\left(t\right)-s\left(t\right)$, the step $a$ in (\ref{eq_pumfdu_1_1}) holds. Due to the steady-state condition (The mean arrival rate is less than or equal to the mean service rate), $\mathbb{E}\left[a_i\left(t+{W}^b\right)\right]\le\mathbb{E}\left[s_i\left(t+{W}^b\right)\right]$, the step $b$ in (\ref{eq_pumfdu_1_1}) holds. And $Y_{t-1}^{delay}$ is $\mathcal{F}_{t-1}$-measurable, $\mathbb{E}\left[Y_{t-1}^{delay}\mid\mathcal{F}_{t-1}\right]=Y_{t-1}^{delay}$, such that $\mathbb{E}\left[Y_t^{delay}|\mathcal{F}_{t-1}\right]\le Y_{t-1}^{delay}$, which implies that $Y_T^{delay}$ is the supermartingale.
Then, based on the Theorem \ref{theorem2}, let $\delta=\mathbb{E}\left[Y_t^{delay}-Y_{t-1}^{delay}\right]$ we have ${O_T}\left[0\right]\le\sum\limits_{i=1}^{{sb}_{seg}}{\frac{2\mathbb{E}\left[\left(Y_t^{delay}-{sb}_{l}^{i}\right)^{-}\right]}{\delta}}$.
Therefore, the maximum rate of ccurrence of $\left\{W\left(t\right)\ge {W}^b\right\}$ is calculated as
\begin{equation}
\label{eq_dbmqos_13_a1_3}
\begin{array}{lll}
{mr}=\frac
{\sum\limits_{i=1}^{{sb}_{seg}}{2\mathbb{E}\left[\left(Y_t^{delay}-{sb}_{l}^{i}\right)^{-}\right]}}{\delta T}.
\end{array}
\end{equation}

\end{sloppypar}
\end{document}